\documentclass[12pt,a4paper]{article}

\usepackage[british]{babel}

\usepackage[a4paper,top=2cm,bottom=2cm,left=2.5cm,right=2.5cm,marginparwidth=1.75cm]{geometry}

\usepackage[style=apa, backend=biber]{biblatex} 
\DeclareLanguageMapping{british}{british-apa} 
\DeclareFieldFormat[article]{volume}{\apanum{#1}} 

\usepackage{amsmath}
\usepackage{graphicx}
\usepackage[colorlinks=true, allcolors=blue]{hyperref}
\usepackage{hyperref}
\usepackage[title]{appendix}
\usepackage{mathrsfs}
\usepackage{amsfonts}
\usepackage{booktabs} 
\usepackage{caption}  
\usepackage{threeparttable} 
\usepackage{algorithm}
\usepackage{algorithmicx}
\usepackage{algpseudocode}
\usepackage{listings}
\usepackage{enumitem}
\usepackage{chngcntr}
\usepackage{booktabs}
\usepackage{lipsum}
\usepackage{subcaption}
\usepackage{authblk}
\usepackage[T1]{fontenc}    
\usepackage{csquotes}       
\usepackage{diagbox}
\usepackage{comment}
\usepackage{bbm}
\usepackage{float}

\usepackage{setspace}
\usepackage{titlesec}
\titleformat{\section} 
  {\normalfont\Large\bfseries}{\thesection.}{1em}{}
  
\usepackage{lineno} 

\rightlinenumbers 

\linenumbers 

\usepackage{float}   
\usepackage{caption} 


\title{Evaluating infectious disease forecasts in a cost-loss situation}

\author[1,*]{Philip Gerlee}
\author[1]{Torbjörn Lundh}
\author[2]{Anna Saxne Jöud}
\author[3]{Henrik Thorén}

\affil[1]{\small Department of Mathematical Sciences, Chalmers University of Technology \& University of Gothenburg, Sweden \texttt{gerlee@chalmers.se}}
\affil[2]{\small Occupational and environmental medicine, Department of Laboratory medicine, Lund University, Lund, Sweden}
\affil[3]{\small Department of research, development, education and innovation, Skåne university hospital, Lund, Sweden}
\affil[4]{\small Department of Philosophy, Lund University, Sweden}
\affil[*]{\small Corresponding author: Philip Gerlee}

\date{}  

\begin{document}
\maketitle

\begin{abstract}
\noindent In order for epidemiological forecasts to be useful for decision-makers the forecasts need to be properly validated and evaluated. Although several metrics fore evaluation have been proposed and used none of them account for the potential costs and losses that the decision-maker faces. We have adapted a decision-theoretic framework to an epidemiological context which assigns a Value Score (VS) to each model by comparing the expected expense of the decision-maker when acting on the model forecast to the expected expense obtained from acting on historical event probabilities. The VS depends on the cost-loss ratio and a positive VS implies added value for the decision-maker whereas a negative VS means that historical event probabilities outperform the model forecasts. We apply this framework to a subset of model forecasts of influenza peak intensity from the FluSight Challenge and show that most models exhibit a positive VS for some range of cost-loss ratios. However, there is no clear relationship between the VS and the original ranking of the model forecasts obtained using a modified log score. This is in part explained by the fact that the VS is sensitive to over- vs. underprediction, which is not the case for standard evaluation metrics. We believe that this type of context-sensitive evaluation will lead to improved utilisation of epidemiological forecasts by decision-makers. 
\end{abstract}


\section{Introduction}
Epidemiological forecasting plays an important role when responding to infectious disease outbreaks. This became evident during the recent COVID-19 pandemic, where forecasts were used to inform decision-makers concerning e.g. the effect of non-pharmaceutical interventions and to aid resource allocation in healthcare \parencite{nixon2022real}. The latter problem was particularly severe during the first phase of the pandemic, when many hospitals responded to forecasts by scaling up the number of regular hospital beds as well as ICU-beds \parencite{lefrant2020national}. These actions were taken at the expense of planned healthcare, such as elective surgeries and other treatments, which were postponed causing potential loss in health \parencite{arsenault2022covid}.

Though less dramatic, forecasting efforts are also made for the seasonal influenza in order to predict the load on healthcare systems in the short-term (1-4 weeks) and the timing and intensity of the peak \parencite{shaman2013real}. In the US these efforts have been channelled through the FluSight Challenge which has been running since 2013 and is operated by the Centers of Disease Control and Surveillance (CDC) \parencite{reich2019collaborative,reich2019accuracy}. FluSight invites researchers to submit weekly real-time forecasts during ongoing influenza seasons. These forecasts are required to be probabilistic, i.e. for each target the forecaster submits a probability distribution of the target variable.

The forecasts were previously focused on weighted Influenza-like-illness (wILI), which is the weighted percentage of outpatient visits for influenza-like illness collected through the U.S. Outpatient Influenza-like Illness Surveillance Network (ILINet) \parencite{CDC_FluView_Overview}. As of the 2021-22 season the target has shifted to lab-confirmed hospitalisations of influenza, which is considered more valuable \parencite{mathis2024evaluation}. 

The forecasts from the FluSight Challenge have been retrospectively evaluated using an exponentiated log score used by the CDC \parencite{reich2019collaborative}. This score is the geometric average of the logarithm of the forecasted probabilities in narrow ranges around the actual outcomes and is a measure both of accuracy and precision. Another common evaluation metric is the Weighted Interval Score (WIS), which averages forecast performance across multiple prediction intervals and rewards accurate and precise forecasts \parencite{bracher2021evaluating}. Also, less sophisticated measures of performance have been used, e.g. Root Mean Squared Error and Mean Absolute Percentage Error, which only take point predictions into account. All these metrics allow for a relative ranking of model forecasts, but are not designed with decision-makers in mind.

One recent attempt to account for policy decisions when evaluating forecasts was made by Gerding et al. who used successful allocation of healthcare resources based on forecasts as means to assign a score \parencite{gerding2024evaluating}. They considered several hospitals where the total amount of healthcare resources are limited and a decision-maker has to decide how to distribute these resources among the hospitals. Model forecasts inform the decision-maker of the expected number of admitted patients to each hospital, and the score is calculated as the total unmet need across all hospitals. This describes a situation that occurred during the COVID-19 pandemic, when resources indeed were limited, but for seasonal outbreaks, such as influenza, decision-makers are faced with a different dilemma.

For decision-makers forecasts of influenza incidence are used for antiviral treatment allocation, to prepare for increases in flu-related hospitalizations and for informing the distribution and placement of health care staff and hospital beds and treatment resources. Beyond healthcare, forecasts can also be used to help guide mitigation strategies, such as non-pharmaceutical interventions, e.g. reducing contacts during times of forecasted high flu activity, and conveying the importance of flu vaccination prior to forecasted increases in flu activity \parencite{CDC_FluForecasting_About}. The decision to act on a forecast of a severe influenza season is associated with certain costs, e.g. advertising costs and costs of purchasing additional antiviral treatments. On the other hand not preparing when a severe season occurs is associated with loss both in money (e.g. increased costs for hospitalisations) and health (both immediate loss in health and long-term effects). 

From the point of view of the decision-maker it is therefore relevant to rank model forecasts with respect to potential costs and losses of the event that is being forecast. Such evaluation frameworks have been developed in meteorology where the value of a (model) forecast can be calculated \parencite{wilks2001skill}. This value is calculated by contrasting the expected expense of the decision-maker if it acts on the prediction of the model with the expected expense of a baseline climatological model, which only makes use of historical data.

In this paper we adapt a simple decision-theoretic framework to an infectious disease setting, where the decision-maker must decide to prepare or not prepare for a severe influenza season at the beginning of the season based on forecasts of peak intensity.
We apply this novel evaluation framework to data from the FluSight Forecasts and show that the value forecasts with respect to the FluSight baseline forecast depends on the ratio of costs for preparations and losses incurred. In addition, we also show that whereas standard evaluaton metrics are symmetric with respect to over- and under-prediction the difference in expected expense of such model forecasts is proportional to the loss, which indeed can be substantial.

\section{Results}
\subsection{Evaluation in a cost-loss framework}
We build on a simple decision-theoretic framework from meteorology first introduced by \cite{murphy1969expected}, where the decision-maker is faced with two potential outcomes: a severe influenza season and a normal one, and is equipped with two actions: to prepare for the severe season or not prepare (see Table \ref{tab:cost_loss}, and \cite{wilks2001skill} for a modern treatment). Preparation in this context refers to campaigns to increase vaccine uptake and increasing vaccination capacity, and comes with a fixed cost $C$. In the event that the decision-maker prepares there are no further costs independent of the outcome, whereas if they decide not to prepare and a severe season occurs then a fixed loss $L$ is incurred. This loss is related to costs for hospitalisations and loss in health of affected patients both in the short and longer term. Note that if the cost exceeds the loss it is always beneficial not to act, and it is therefore customary to assume $0<C<L$ or equivalently that the cost-loss-ratio satisfies $0 <C/L < 1$.

\begin{table}[htbp]
\centering
\caption{Cost–loss decision table for preparing for a severe influenza season. The entry in each cell is the cost incurred by the decision-maker under the corresponding action-outcome pair.}
\label{tab:cost_loss}
\begin{tabular}{lcc}
\toprule
& \multicolumn{2}{c}{\textbf{Outcome}} \\
\cmidrule(lr){2-3}
\textbf{Action} & Severe season & Normal season \\
\midrule
Prepare & $C$ & $C$ \\
Do not prepare & $L$ & $0$ \\
\bottomrule
\end{tabular}
\end{table}

If the decision-maker tries to minimise their expected expense, and they believe that a severe season will occur with probability $p$, then they should act whenever the expense of acting ($C$) is less than the expected expense of not acting ($pL$). In other words if $C < pL$, or equivalently  $p > C/L$. Now such a probability can be obtained using a forecasting model, which makes use of data up until some specific date, or it can be estimated by only considering data from previous seasons (often referred to as the climatological probability in weather forecasting), where the event probability $p$ is estimated as the fraction of seasons that in the past were severe. We refer to this as the baseline event probability $p_b$. By averaging the expense incurred by acting on the forecasts from the model ($E_f$) and the baseline ($E_b$), over several seasons we can calculate a Value Score \parencite{wilks2001skill}:
\begin{equation}\label{eq:VS}
\mbox{\mbox{VS}} = \frac{E_b-E_f}{E_b - E_p},
\end{equation}
where $E_p$ is the expected expense when relying on a perfect or oracle forecast, where actions are taken precisely on those season when required. Since no real forecast model can outperform the oracle it normalises the Value Score such that $-\infty < \mbox{VS}<1$. Note that since the actions taken when following the model and baseline forecasts depend on the cost-loss-ratio, the Value Score depends on $C/L$ and is often depicted in graph where $C/L$ ranges from 0 to 1. A $\mbox{VS}>0$ for a given $C/L$ implies that a decision-maker who is trying to minimise their expected expense is to prefer the model forecast over the baseline, whereas $\mbox{VS}<0$ suggests that acting on the baseline model is preferable. A $\mbox{VS}=1$ means that the model forecast is on par with a perfect forecast.

\subsection{Evaluation of influenza peak intensity forecasts}
\begin{figure}[!htb]
\centering
\includegraphics[width=1\linewidth]{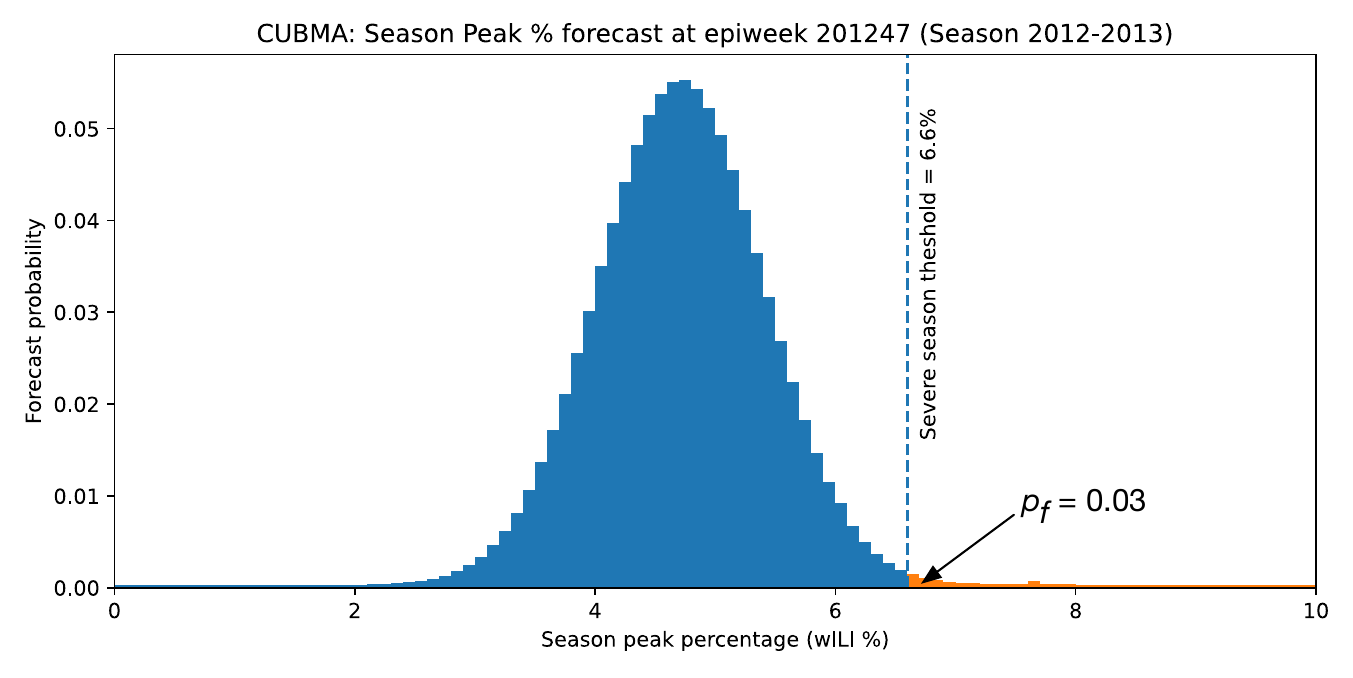}
\caption{\label{fig:pf}An example of a probabilistic forecast of the wILI peak intensity made at epiweek 47 when wILI first crosses the seasonal baseline. The dashed line corresponds to the threshold value for a severe season and the orange bars show the probability of a severe season, which in this case equals $p_f=0.03$.}
\end{figure}

In the context of the FluSight Challenge we assume that the decision-maker acts on forecasts of peak wILI that are made during the first week of the season when wILI crosses the CDC baseline \parencite{biggerstaff2018systematic}, which typically occurs during the autumn. A severe season is defined as peak wILI exceeding the $TI_{90}$ threshold, as defined in \parencite{biggerstaff2018systematic}, which equals 6.6 \%. We use the probabilistic model forecast of peak intensity to calculate the event probability $p_f$, i.e. the probability that peak intensity exceeds 6.6 \%. An example of a probabilistic forecast of the wILI peak intensity from the \texttt{CU-BMA} model is shown in fig. \ref{fig:pf}.

\begin{figure}[!htb]
\centering
\includegraphics[width=1\linewidth]{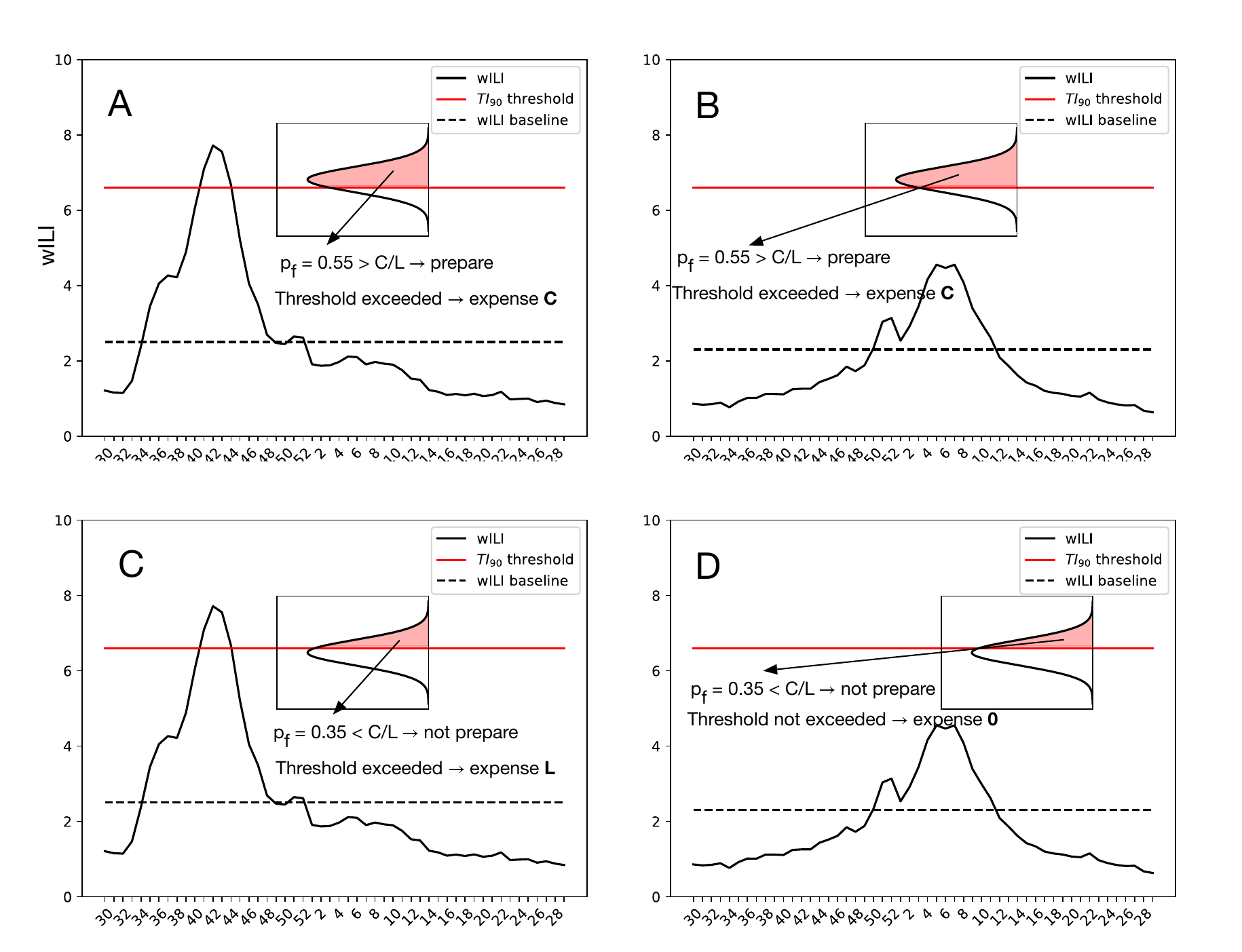}
\caption{\label{fig:outcomes}A schematic of the cost-loss framework as applied to the FluSight Challenge data. Each panel shows the weekly wILI over an influenza season, and the inset shows the forecasted probability distribution of the peak intensity. At the first week when wILI exceeds the baseline (dashed line) the forecaster issues a probabilistic prediction of the wILI peak intensity. If the forecasted probability that peak intensity exceeds the severity thresholds $TI_{90}$ (red solid line) is larger than the cost-loss threshold, which in this example is assumed to be $C/L=0.4$, the decision-maker prepares for a severe season. In this case the expense is given by $C$ independent if the peak intensity exceeds the threshold or not (panel A and B). If the forecasted probability is less than $C/L$ no action is taken, and in the absence of a severe season the expense is zero (panel D). However, if the threshold is exceeded a loss $L$ is incurred (panel C).}
\end{figure}

According to the above reasoning the decision-maker prepares for a severe season if the forecasted event probability satisfies $p_f > C/L$. Four outcomes are possible depending on the relationship between $p_f$ and $C/L$, and if a severe season occurs or not. Each outcome is associated with the expenses shown in table \ref{tab:cost_loss}. A schematic of all the possible outcomes can be seen in fig.\ \ref{fig:outcomes}. 

\begin{figure}[!htb]
\centering
\includegraphics[width=1\linewidth]{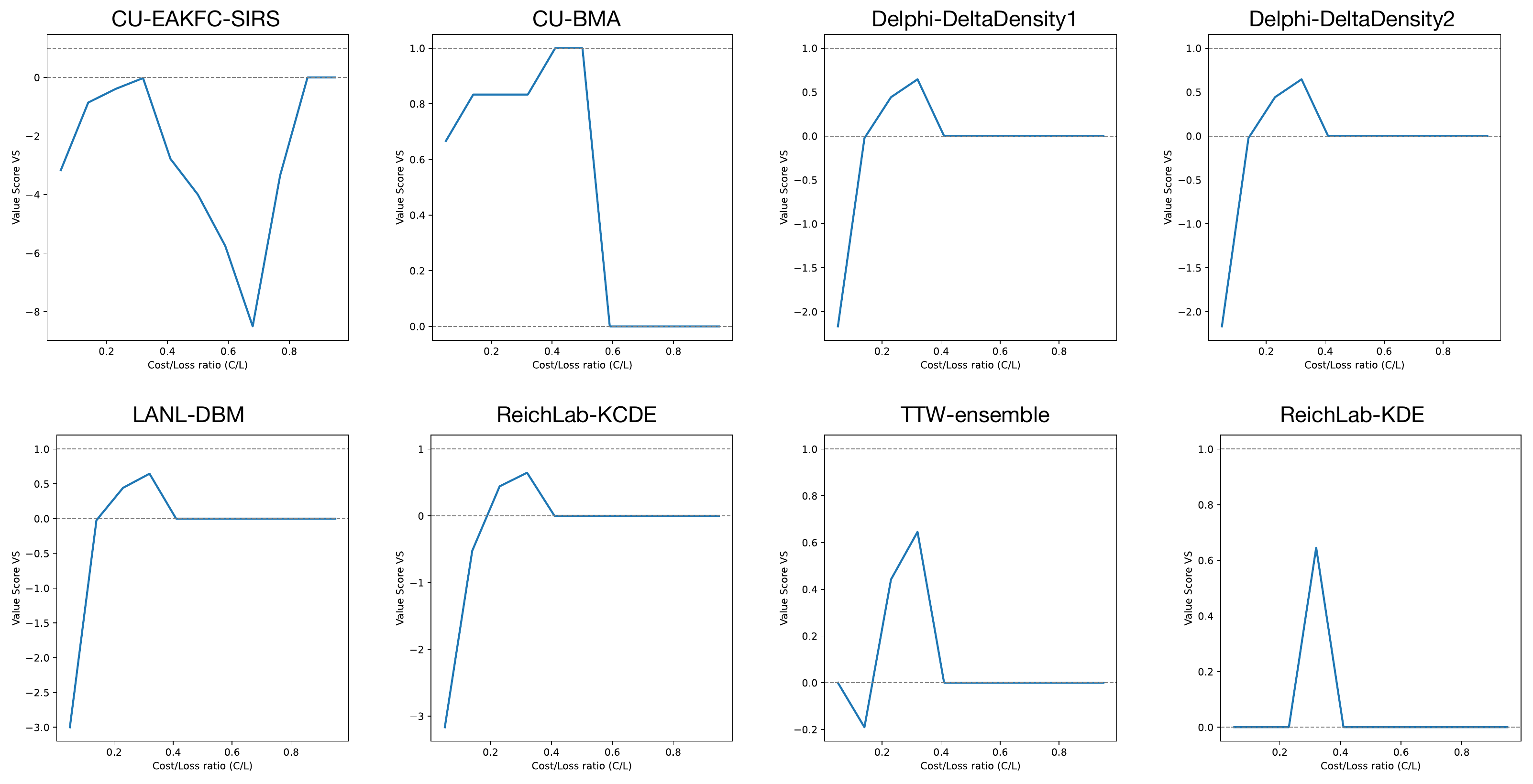}
\caption{\label{fig:VS}Value Score of forecasts from the FluSight Challenge including the Target-Type Weights (TTW) ensemble and the ReichLab-KDE, which is a historical baseline model. The Value Score is calculated relative to a historical baseline event probability of $p_b=0.32$. The \mbox{VS} is plotted against the cost-loss ratio $C/L$. A $\mbox{VS}>0$ for a given $C/L$ implies that a decision-maker who is trying to minimise their expected expense is to prefer the model forecast over the baseline, whereas $\mbox{VS}<0$ suggests that acting on the baseline model is preferable. }
\end{figure}

We consider the influenza seasons from 2011/12 to 2017/18 for which model forecasts are available via the FluSight Challenge GitHub-repo \parencite{reich2019collaborative}. Baseline probabilities are calculated as the fraction of seasons which have a peak intensity larger than the $TI_{90}$ threshold. For illustrative pruposes we consider the models which are included in the FluSight Network Target-Type Weights for seasonal targets \parencite{reich2019accuracy}, and also include the Target-Type Weights ensemble forecasts and ReichLab-KDE, which is a historical baseline model that does not make use of the data from the current season. The resulting Value Score-plots are shown in figure \ref{fig:VS} where a positive $\mbox{VS}$ implies added value for the decision-maker, whereas a negative value means that acting on the baseline probabilities provide lower expenses for the decision-maker. We note that the $\mbox{VS}$-plots are heterogenous falling roughly into four classes: mostly negative values (\texttt{CU-EAKFC-SIRS}), mostly positive values (\texttt{CU-BMA}), large negative for small $C/L$ followed by a positive region (\texttt{Delphi-Density1}, \texttt{Delphi-Density2}, \texttt{LANL-DBM} and \texttt{ReichLab-KCDE}) and close to zero for small $C/L$ and followed by a positive region (\texttt{TTW-ensemble} and \texttt{ReichLab-KDE}). Calculating the Value Score based only on preceding seasons yields qualitatively similar results (see Supplementary Figure 1).

Although the \mbox{VS} is a function of $C/L$ it is possible to summarise it by averaging it over a range of $C/L$ that the decision-maker considers reasonable. At this point we do not have such a range (see Discussion for an attempt to list involved costs and losses) and we therefore average over the entire $C/L$-range in figure \ref{fig:VS}, which equals $0.05 < C/L < 0.95$. The results of this averaging is shown in table \ref{tab:mVS}.

\begin{table}[ht]
\centering
\caption{Average Value Score of the considered models over the range $0.05 < C/L < 0.95$.}
\label{tab:mVS}
\begin{tabular}{lr}
\hline
\textbf{Model} & \textbf{Average \mbox{VS}} \\
\hline
CU-BMA        &  0.47 \\
ReichLab-KDE           &  0.06 \\
Delphi-Density1       & -0.10 \\
Delphi-Density2       & -0.10 \\
LANL-DBM      & -0.18 \\
ReichLab-KCDE          & -0.24 \\
CU-EAKFC-SIRS & -2.60 \\
\hline
\end{tabular}
\end{table}

In order to get a better understanding of how the \mbox{VS} is calculated we show the forecasted probabilities of severe seasons in table \ref{tab:details} for \texttt{CU-BMA}. From the table it can be seen that the model forecasts very low probabilities in 5 out of 6 seasons when $TI_{90}$ was not exceeded and a large probability for the 2017 season when it was exceeded. For a $C/L=0.5$ the forecasted probabilities are such that a rational decision-maker should act only during the season when the threshold was exceeded, which results in a maximal $\mbox{VS}=1$ for that cost-loss ratio (see fig.\ \ref{fig:VS}). We also see that for $C/L>0.545$ a rational decision-maker would never prepare during the considered seasons (since all $p_f$'s are less than 0.545). The same holds for a decision-maker following the baseline probability ($p_b=0.32$), and thus they have the same expected expense and therefore $\mbox{VS}=0$ for \texttt{CU-BMA} when $C/L>0.545$.

\begin{table}[ht]
\centering
\caption{Details of the \texttt{CU-BMA} forecasts of peak intensity for the different seasons. 2011 is excluded since the wILI baseline was never crossed that season.}
\label{tab:details}
\begin{tabular}{rrrrr}
\hline
\textbf{Year} & \textbf{Week of forecast} & \textbf{Event} & $\mathbf{p_{\text{f}}}$ & $\mathbf{p_b}$ \\
\hline
2010 & 51 & No & 0.020 & 0.320 \\
2011 & - & - & - & - \\
2012 & 47 & No & 0.026 & 0.320 \\
2013 & 48 & No & 0.023 & 0.320 \\
2014 & 47 & No & 0.093 & 0.320 \\
2015 & 51 & No & 0.385 & 0.320 \\
2016 & 50 & No & 0.018 & 0.320 \\
2017 & 47 & Yes & 0.545 & 0.320 \\
\hline
\end{tabular}
\end{table}

\subsection{Value Score depends on time of forecast}
\begin{figure}[htb]
\centering
\includegraphics[width=1\linewidth]{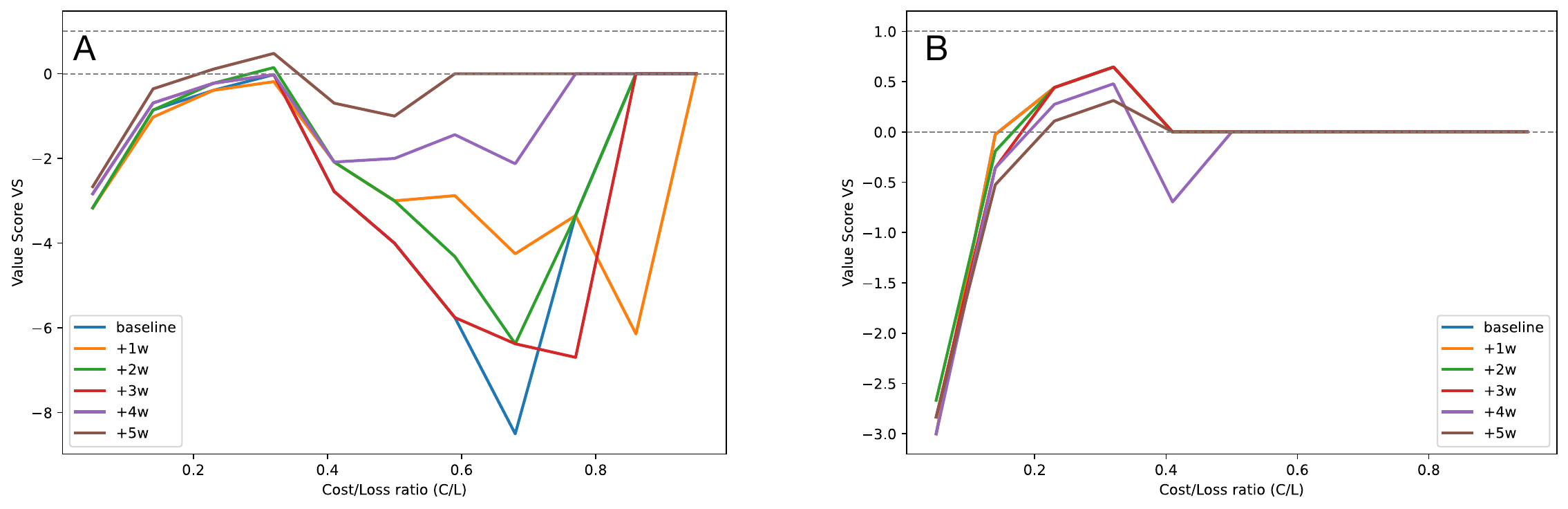}
\caption{\label{fig:horizon}The Value Score of the A) \texttt{CU-EAKFC-SIRS}-model and B) the \texttt{LANL-DBM}-model for forecasts of wILI peak intensity exceeding $TI_{90}$ made at season onset and 1-5 weeks later.}
\end{figure}

Thus far we have evaluated forecasts made at the start of the influenza season when the wILI crosses the baseline. However, we expect that the value of a forecast of the peak intensity, as compared to the historical baseline forecast, should depend on when the forecast is made. To investigate this we compare the \mbox{VS} of forecast made at season onset with those made 1-5 weeks later. The result of this comparison can be seen in fig. \ref{fig:horizon} which shows that the \mbox{VS} for the \texttt{CU-EAKFC-SIRS}-model (fig. \ref{fig:horizon}A) increases for later forecasts, whereas the \texttt{LANL-DBM}-model (fig. \ref{fig:horizon}B) performs worse with respect to \mbox{VS} later in the season. 

\subsection{Relative Value Score of two models}
\begin{figure}[htb]
\centering
\includegraphics[width=1\linewidth]{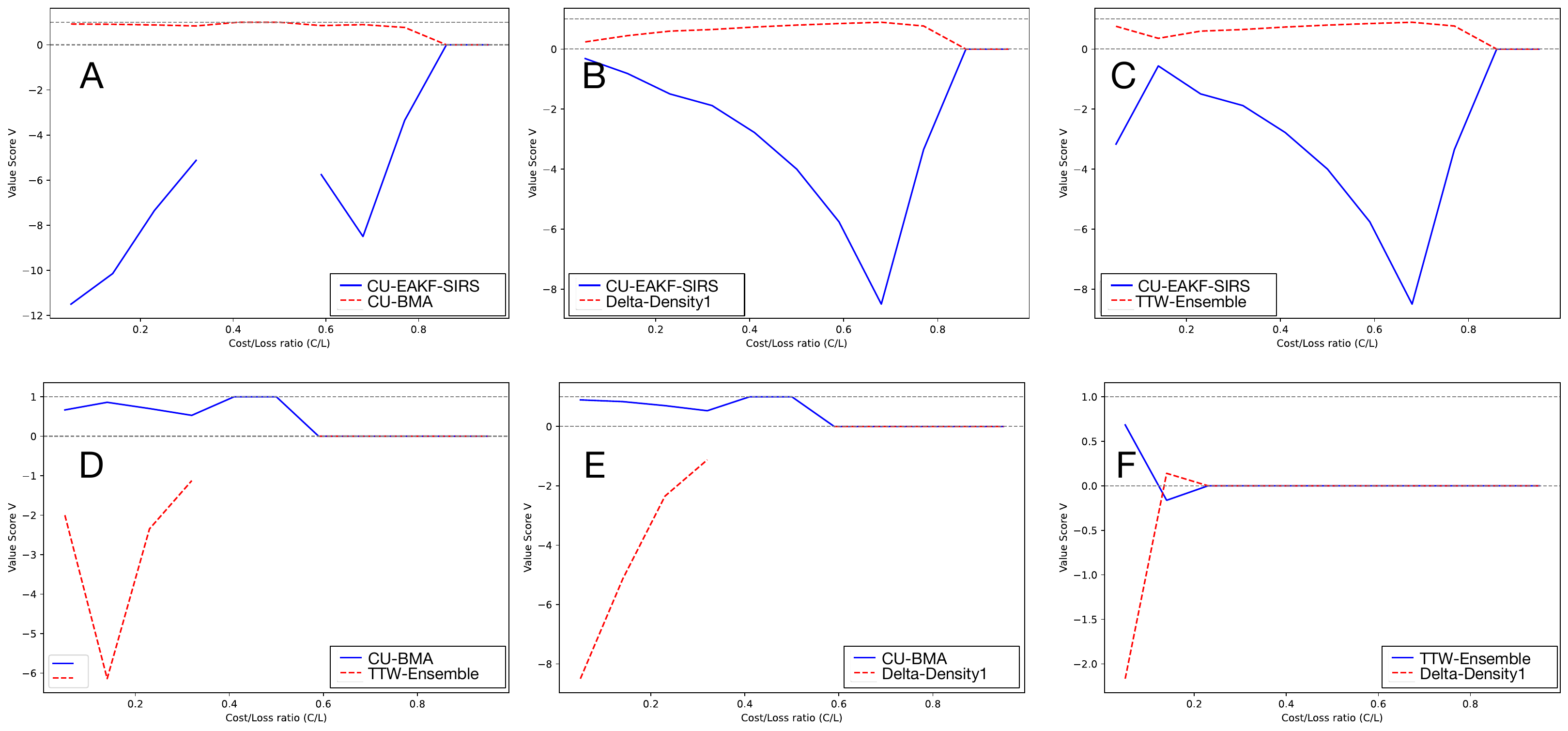}
\caption{\label{fig:VSab}The Value Score of model A relative to B for \texttt{CU-EAKFC-SIRS}, \texttt{CU-BMA}, \texttt{Delphi-Density1} and \texttt{TTW-ensemble}. Each line (red dashed or solid blue) corresponds to that model relative to the other model in that panel. Note that the value score of model A relative to B is undefined when model B produces forecasts on par with the perfect forecast.}
\end{figure}

The Value Score as defined in \eqref{eq:VS} compares the expected expense of acting on a given model forecast ($E_f$) to that of a baseline probability ($E_b$). It could also be of interest to compare the value of a forecast from model A to that of model B. To do this we define the Value Score of model A relative to B as:
\begin{equation}\label{eq:VSab}
\mbox{VS}(A,B) = \frac{E_{B}-E_A}{E_B - E_p},
\end{equation}
where $E_{A,B}$ is the expected expense when acting on forecast A or B, and $E_p$ again is the minimal expense obtained from acting on a perfect forecast. In general we have that this metric is not symmetric, i.e. $\mbox{VS}(A,B) \neq \mbox{VS}(B,A)$. To illustrate this relative Value Score we calculated it for models in the four classes of $\mbox{VS}$-plots discussed above. The result can be seen in fig. \ref{fig:VSab} which shows the pair-wise Value Score for the \texttt{CU-EAKFC-SIRS}, \texttt{CU-BMA}, 
\texttt{Delphi-Density1} and \texttt{TTW-ensemble}. From these plots we can conclude that the other models are preferred over \texttt{CU-EAKFC-SIRS} for all $C/L<0.8$ (fig.\ \ref{fig:VSab}A-C) and that \texttt{CU-BMA} is preferred over the remaining two models for $C/L < 0.6$ (fig.\ \ref{fig:VSab}D-E). Lastly, \texttt{Delphi-Density1} and \texttt{TTW-ensemble} are equivalent except for small values of $C/L$ (fig.\ \ref{fig:VSab}F). Note that the VS of model A relative to B is undefined when model B produces forecasts on par with the perfect forecast, since the denominator of \eqref{eq:VSab} in that case equals zero.

\subsection{The Value Score is not symmetric}
Standard metrics for evaluation of epidemic forecasts such as Weighted Interval Score (WIS) and the Logarithmic Score (LS) do not account for the consequences of over- and underprediction. Indeed it can be shown that if the target value has a Gaussian distribution with mean $\mu$ and the forecasts are also Gaussian, then two forecast that over- and underpredict in equal amounts obtain the same expected WIS and LS (see Supplementary Material for a detailed proof). The expected expense (and therefore the Value Score) is, due to the asymmetry of the cost-loss framework, not symmetric with respect to under- and overprediction. To illustrate this fact consider two models with predictive distributions that are Gaussian with mean $\mu \pm a$ and common variance $\sigma_f^2$. Further assume that an event occurs when the target value exceeds a fixed threshold $T$, and denote the expected expense when acting on these models as $E_\pm$ respectively. Then it can be shown that when the two models recommend different actions (a precise criteria for when this occurs is given in the supplement), the difference in expected expense is given by
\begin{equation}\label{eq:diff}
\Delta E = E_+ - E_- = L(C/L-p),
\end{equation}
where $p$ is the true event probability. This implies that when $C/L$ is larger than the true event probability, the model that overpredicts the target variable has a larger expected expense, and the difference is proportional to the loss $L$. Vice versa, if $C/L<p$ the over-predicting model achieves a lower expected expense. A detailed proof is provided in the Supplement. 

\section{Discussion}
We have shown that a cost-loss framework initially formulated for meteorological forecasts can be adapted to infectious disease epidemiology. In contrast to standard evaluation metrics in epidemiology the Value Score is context-dependent and provides a metric which takes into account the economic and health economic consequences that a decision-maker needs to consider. The Value Score was calculated for forecasts of peak intensity of wILI from the FluSight Challenge, which showed that 5 out 6 component models in the seasonal ensemble had positive value for a decision-maker at some cost-loss ratio. The model with negative Value Score (\texttt{CU-EAKFC-SIRS}) showed improvement in $\mbox{VS}$ when forecasts were made further into the season. We also devised a relative Value Score which makes it possible for a decision-maker to decide which of two models provide most value depending on the cost-loss ratio. Lastly, we showed that the $\mbox{VS}$ is sensitive to over- vs. underprediction in contrast to standard evaluation metrics. 

The Value Score depends on the costs and losses in the decision framework (see table \ref{tab:cost_loss}), and is therefore visualised as graph with $C/L$ on the x-axis. Finding an exact value for $C/L$ in the context of a severe influenza season is not possible, but it should be possible for decision-makers to estimate cost and losses involved. The cost of preparing for a severe season include the costs for communication campaigns in pharmacies, TV and social media in order to increase vaccine uptake \parencite{kansagra2012cost}, the costs of additional vaccine doses and administration of vaccine by healthcare staff \parencite{walsh2010economic}. The losses incurred during a severe season include the cost of hospitalised influenza patients \parencite{hu2024economic}, lost productivity and wages due to sick leave \parencite{de2022economic} and loss in health of those infected with influenza, in particular those that are hospitalised which experience both short and longer term losses in health \parencite{sandmann2022health}. Converting the losses in health into monetary terms could be achieved with a health economic analysis \parencite{iino2022estimating}, and would make it possible to at least obtain an estimate of $C/L$.

The Value Score cannot be compared in a straight-forward way to other evaluation metric since it depends on the cost-loss ratio. In the absence of a estimate or reasonable range for $C/L$ it is possible to average the across the entire range, as was done in table \ref{tab:mVS}. In this comparison \texttt{CU-BMA} comes out as the model with the largest value followed by the \texttt{ReichLab-KDE}. The latter model serves as the historical baseline model in the FluSight Challenge. All other models we consider (including the ensemble) have negative average \mbox{VS}, which is due to the large negative \mbox{VS} they exhibit for small $C/L$. This corresponds to a decision situation when the loss incurred by not acting during a severe season is many times larger than the costs of preparing. In this setting the historical baseline model with an event probability of $p_b=0.32$ is preferred over all models except the \texttt{CU-BMA} and the \texttt{ReichLab-KDE}. In the original evaluation of the FluSight Challenge, which made use of a modified log score, the \texttt{ReichLab-KDE} performed worst among the models considered here with respect to seasonal target, while \texttt{Delphi-DeltaDensity2} was the top-performing model closely followed by \texttt{Delphi-DeltaDensity1}, \texttt{LANL-DBM} and \texttt{ReichLab-KCDE}. Surprisingly the \texttt{CU-BMA} performed second to worst with respect to the log score, but here shows the highest average \mbox{VS}. This points to the fact that there is no simple relationship between the log score and the \mbox{VS}. A similar observation was made by Gerding et al. when they considered a scoring rule based on allocation of resources and compared it to the weighted interval score \parencite{gerding2024evaluating}.

The improvement in \mbox{VS} for forecasts made later in the season for \texttt{CU-EAKFC-SIRS} (fig. \ref{fig:horizon}) agrees with what was seen in \parencite{reich2019accuracy}, where a slight improvement in score is seen closer to the peak. A similar pattern in score is seen for \texttt{LANL-DBM}, but this is not reflected in an increased \mbox{VS} as the season proceeds. Again this highlights the difficulty of relating the two measures. 

Another perspective we may assume on this framework is in terms of what philosophers have discussed under the label ``inductive risk'' \parencite{douglas2009science, rudner1953scientist}. In short inductive risks concern risks of drawing the wrong conclusions when action hinges on that conclusion. The cases described above are perfect illustrations. Wrongly accepting the prediction that there will be no severe flu season activates considerations of inductive risks as accepting this particular prediction means that the decision-maker will not prepare and hence must take the losses associated with an unmitigated severe flu season. To care about inductive risks involves more than simply to care about being wrong (which might be labelled epistemic risk plain and simple) precisely because inductive risk considerations integrate further harms, a difference that is highlighted when those harms are asymmetrically distributed over possible errors. The idea is that the risk of harmful errors should be minimised.
The argument is useful in this context as it tells us something about the benefits of VS compared to other measures, and the conditions under which it may be useful. Under situations when (a) costs and losses can be reliably estimated (or are broadly agreed upon), and hence the distribution of harms can be sufficiently well established, (b) those costs and losses are asymmetric in the way outlined above, inductive risk considerations should guide how the performance of models is evaluated as the central aim is harm avoidance and not mere accuracy. This is exactly what this framework does.

This study is a first attempt to evaluate epidemic forecasts in a cost-loss framework, and as such it has several limitations. Firstly, we consider a strongly simplified decision framework with binary outcomes (severe vs. not severe) and actions (prepare vs. not prepare). A natural extensions would be to consider a framework where the loss is proportional to the severity, similar to what has been proposed by Lee \& Lee in a weather forecast context \parencite{lee2007economic}, which also allows for a varied response that depends on the forecasted event probability. Secondly, we only consider a single event (a severe season). It is also possible to consider other events, e.g. the peak week occurring early (before some fixed week) or weekly events, such as ``the wILI in 4 weeks time will be 50\% higher than the current week''. Although the event we have investigated here is relevant for decision-makers it has the drawback of containing few data points. In total we considered 8 seasons of wILI measurements, but since the wILI baseline threshold was never crossed during the 2011/2012 season, the Value Score calculations are based on only 7 events. In future work it would be interesting to analyse a larger dataset, e.g. by considering weekly targets, which would make it possible to investigate the relationship between \mbox{VS} and other evaluation metrics in more detail.

Despite these limitations we believe that the Value Score of an epidemic forecasting model could be a useful tool for decision-makers when evaluating the utility of different models in a situation where the cost-loss ratio can be estimated to lie within a certain range.

\section*{Declarations}
\subsection*{Author contributions}
The study was conceptualised by P.G., T.L., A.J.S and H.T. The method was developed by all authors and implemented by P.G. The manuscript text was written by P.G. and H.T. and edited and reviewed by all authors. All figures were prepared by P.G.

\subsection*{Data Availability}
All code used in this study is available at \href{https://github.com/philipgerlee/Evaluating-infectious-disease-forecasts-in-a-cost-loss-situation}{GitHub}. The data used in this study is available at the \href{https://github.com/FluSightNetwork/cdc-flusight-ensemble/tree/first-papers/model-forecasts}{FluSight Github repo}.

\subsection*{Funding information}
This work was carried out with funding from Vetenskapsrådet grant number 2022-06368.

\subsection*{Competing interest}
The authors declare no competing interests.

\clearpage
\printbibliography
\clearpage

\section{Supplementary methods}
\subsection{Preliminaries}
We consider a simplified prediction situation where the target variable $X$ has a normal distribution with mean $\mu$ and variance $\sigma^2$. Forecasts are made with two models that produce probabilistic predictions that are also Gaussian. One model $F^{-}$ underpredicts the target, and is distributed according to $N(\mu-a,\sigma_f^2)$, whereas the other model $F^+$ overpredicts the target in equal amount and is distributed according to $N(\mu+a,\sigma_f^2)$. We now proceed to calculate the expected score of these models with respect to the Logarithmic Score and the Weighted Interval Score.

\subsection{Logarithmic score}
The logarithmic score is defined as $L(F,x)=\ln p(x)$, where $p(x)$ is the probability density assigned to outcome $x$ by the model $F$. In order to calculate the expected logarithmic score we need to take the expectation with respect to the outcome $X$, which is a random variable, with a Gaussian distribution that we denote $f(x)$:
\begin{align*}
\mathbb{E}[L(F^-,X)] &= \int_{-\infty}^{\infty} f(x) \ln \left( \frac{e^{-(x-(\mu-a))^2/2\sigma_f^2}}{\sqrt{2\pi \sigma_f^2}} \right) dx = \\
 &=  \int_{-\infty}^{\infty} f(x) \left( - \frac{(x-(\mu-a))^2}{2\sigma_f^2} - \ln \sqrt{2\pi \sigma_f^2} \right) dx = \\
 &=  - \ln \sqrt{2\pi \sigma_f^2} - \frac{1}{2 \sigma_f^2} \int_{-\infty}^{\infty} f(x) \left(x-(\mu-a) \right)^2 dx = \\
 &=  - \ln \sqrt{2\pi \sigma_f^2} - \frac{1}{2 \sigma_f^2} \int_{-\infty}^{\infty} f(x) \left(x^2+\mu^2 +a^2 -2\mu a -2x\mu + 2xa \right) dx =  \\
 &=  - \ln \sqrt{2\pi \sigma_f^2} - \frac{1}{2 \sigma_f^2} \left(\sigma^2+\mu^2 + \mu^2 +a^2 - 2\mu a -2\mu^2 + 2\mu a \right) = \\
 &=  - \ln \sqrt{2\pi \sigma_f^2} - \frac{1}{2 \sigma_f^2} \left(\sigma^2 + a^2 \right). 
\end{align*}
The analogue calculation for $F_+$ yields:
\begin{align*}
\mathbb{E}[L(F^+,X)] &= \int_{-\infty}^{\infty} f(x) \ln \left( \frac{e^{-(x-(\mu+a))^2/2\sigma_f^2}}{\sqrt{2\pi \sigma_f^2}} \right) dx = \\
 &=  \int_{-\infty}^{\infty} f(x) \left( - \frac{(x-(\mu+a))^2}{2\sigma_f^2} - \ln \sqrt{2\pi \sigma_f^2} \right) dx = \\
 &=  - \ln \sqrt{2\pi \sigma_f^2} -\frac{1}{2 \sigma_f^2} \int_{-\infty}^{\infty} f(x) \left(x-(\mu+a) \right)^2 dx = \\
 &=  - \ln \sqrt{2\pi \sigma_f^2} - \frac{1}{2 \sigma_f^2} \int_{-\infty}^{\infty} f(x) \left(x^2+\mu^2 +a^2 + 2\mu a -2x\mu - 2xa \right) dx =  \\
 &=  - \ln \sqrt{2\pi \sigma_f^2} - \frac{1}{2 \sigma_f^2} \left(\sigma^2+\mu^2 + \mu^2 +a^2 + 2\mu a -2\mu^2 - 2\mu a \right) = \\
 &=  - \ln \sqrt{2\pi \sigma_f^2} - \frac{1}{2 \sigma_f^2} \left(\sigma^2 + a^2 \right) =\\
 &= \mathbb{E}[L(F^-,X)].
\end{align*}
Thus, we have shown that logarithmic score yields the same score for $F_-$ and $F_+$ that under- and overpredict in equal amounts.

\subsection{Weighted Interval Score (WIS)}
Let $\alpha \in ]0, 1[$, and let $\hat{x}$ be the point prediction of the model, given by the mean of the predictive distribution and $x$ the outcome. Denote the prediction interval at significance level $\alpha$ of the model by $[l_\alpha, u_\alpha]$. The Interval Score at significance level $\alpha$ is defined as
\begin{equation*}
\mbox{IS}_\alpha([l_\alpha, u_\alpha],  x) = \frac{2}{\alpha} \left(\mathbbm{1}_{\{x<l_\alpha\}} (l_\alpha-x) + \mathbbm{1}_{\{x>u_\alpha\}} (x-u_\alpha) + (u_\alpha-l_\alpha)\right).    
\end{equation*}
This metric consists of three terms: a term of overprediction that punishes a model with a prediction interval at level $\alpha$ which is above the real value, a term of underprediction that punishes a model whose prediction interval is under the real value, and a term of range, that punishes too wide prediction intervals. \\
Let $(\alpha_k)_{k \in \{1, \dots , K\}} \in ] 0 , 1 [ ^K $ be a sequence of significance levels. The WIS is now defined as 

\begin{equation}\label{eq:WIS}
\mbox{WIS}(F, \hat{x},x) = w_0 |x-\hat{x}| + \sum_{k=1}^{K} w_k \mbox{IS}_{\alpha_k}([l_{\alpha_k}, u_{\alpha_k}], x),    
\end{equation}
with weights $(w_k)_{k \in \{0, ... , K\}} \in \mathbb{R}_+ ^K $ chosen by the user.

For the two Gaussian forecast models defined above the prediction intervals at level $\alpha$ are given by
\begin{align*}
l_\alpha^\pm &= \mu \pm a -c \\
u_\alpha^\pm &= \mu \pm a  + c,
\end{align*}
where $\pm$ refers to the model that over- or underpredicts, and $c=z_{1-\alpha/2}\sigma_f$. Here $z_{1-\alpha/2}$ the standard normal quantile at level ${1-\alpha/2}$. Let us now denote expected values of the terms corresponding to the upper and lower bounds of the prediction interval in $\mbox{IS}_\alpha$ by
\begin{align*}
T^\pm &= \mathbb{E}[\frac{2}{\alpha}(l_\alpha^\pm -X) \mathbbm{1}_{\{X<l_\alpha^\pm \}}], \\
S^\pm &= \mathbb{E}[\frac{2}{\alpha}(X-u_\alpha^\pm) \mathbbm{1}_{\{X>u_\alpha^\pm \}}].
\end{align*}
Now define $X' = 2\mu-X$, which mirrors $X$ around the mean $\mu$ of the target distribution. Now we have $X < l_\alpha^+ \iff 2\mu - X' < l_\alpha^+ \iff X' > 2\mu - l_\alpha^+$. But $2\mu - l_\alpha^+ = 2\mu - (\mu + a -c) = \mu -a +c = u_\alpha^-$. And thus, $X < l_\alpha^+ \iff X' >u_\alpha^-$. We also have that $l_\alpha^+ - X = X'-u_\alpha^-$.

Due to the symmetry of the normal distribution about $\mu$, the two random variables $X$ and $X'$ have the same probability distribution. Therefore
\begin{align*}
T^+ &= \mathbb{E}[\frac{2}{\alpha}(l_\alpha^+-X) \mathbbm{1}_{\{X<l_\alpha^+ \}}] = \\
&= \mathbb{E}[\frac{2}{\alpha}(X'-u_\alpha^-) \mathbbm{1}_{\{X'>u_\alpha^- \}}] = \\
&= \mathbb{E}[\frac{2}{\alpha}(X-u_\alpha^-) \mathbbm{1}_{\{X>u_\alpha^- \}}] = S^-.
\end{align*}
By the same argument one can show that $S^+ = T^-$. Lastly, we note that the third term of $\mbox{IS}_\alpha$ corresponds to the width of the prediction interval which is identical for the two models. Thus we have shown that 
\begin{align*}
\mathbb{E}[\mbox{IS}_\alpha([l_\alpha^+, u_\alpha^+],  X)]= T^+ + S^+ + 2c = S^- + T^- + 2c = \mathbb{E}[\mbox{IS}_\alpha([l_\alpha^-, u_\alpha^-],  X)]
\end{align*}
To conclude, we note that the term in WIS corresponding to the absolute error of the point prediction satisfies $\mathbb{E}[|X-(\mu+a)|] = \mathbb{E}[|X-(\mu-a)|]$ (again due to the symmetry about $\mu$). Therefore we can deduce that $\mathbb{E}[\mbox{WIS}(F^-, \hat{x},X)]=\mathbb{E}[\mbox{WIS}(F^+, \hat{x},X)$].

\subsection{Expected expense}
We will now investigate how the two forecast models perform with respect to the expected expense in the cost-loss situation we consider. In particular, we will calculate the difference in expected expense for the two models.

With the same target distribution and two forecast models as above we define a severe season as the target variable exceeding some predetermined threshold $T>0$, i.e. when $X>T$. For brevity we denote the cost-loss ratio by $C/L=\tau$. The true event probability can then be written as
 \begin{align*}
p=\mbox{Pr}(X > T) = \phi \left( \frac{\mu- T }{\sigma}\right),
\end{align*}
where $\phi(\cdot)$ is the cumulative distribution function of the standard normal distribution. A rational decision-maker that acts according to model forecast $F^\pm$ should prepare for a severe season whenever the model-based event probability $q^\pm > \tau$. This is equivalent to
\begin{align*}
\phi \left( \frac{\mu \pm a - T }{\sigma_f}\right) > \tau \iff \\
\frac{\mu \pm a - T }{\sigma_f} > \phi^{-1}(\tau) \iff \\
\mu \pm a > T+ \sigma_f \phi^{-1}(\tau) = \rho.
\end{align*}
This implies that the expected expense of acting on model forecast $F^+$ is given by
$$\mathcal{E} (F^+) = \begin{cases}
  C  & \text{ if } \mu+a>\rho, \\
  pL & \text{ otherwise.}
\end{cases}$$
Similarly we get
$$\mathcal{E} (F^-) = \begin{cases}
  C  & \text{ if } \mu-a>\rho, \\
  pL & \text{ otherwise.}
\end{cases}$$
Now, if both $\mu \pm a \geq \rho$ or $\mu \pm a \leq \rho$ we get the same action and hence $\Delta \mathcal{E} = \mathcal{E} (F^+)- \mathcal{E} (F^-) = 0$. 

Now consider the case $\mu - a < \rho < \mu + a \iff |\mu - a| < \rho$. In this case we get $\mathcal{E} (F^-) = pL$ and $\mathcal{E} (F^+)=C$, which implies that $\Delta \mathcal{E} = C-pL = L(C/L-p)=L(\tau-p)$.
\section{Supplementary figures}
\begin{figure}[H]
\centering
\includegraphics[width=1\linewidth]{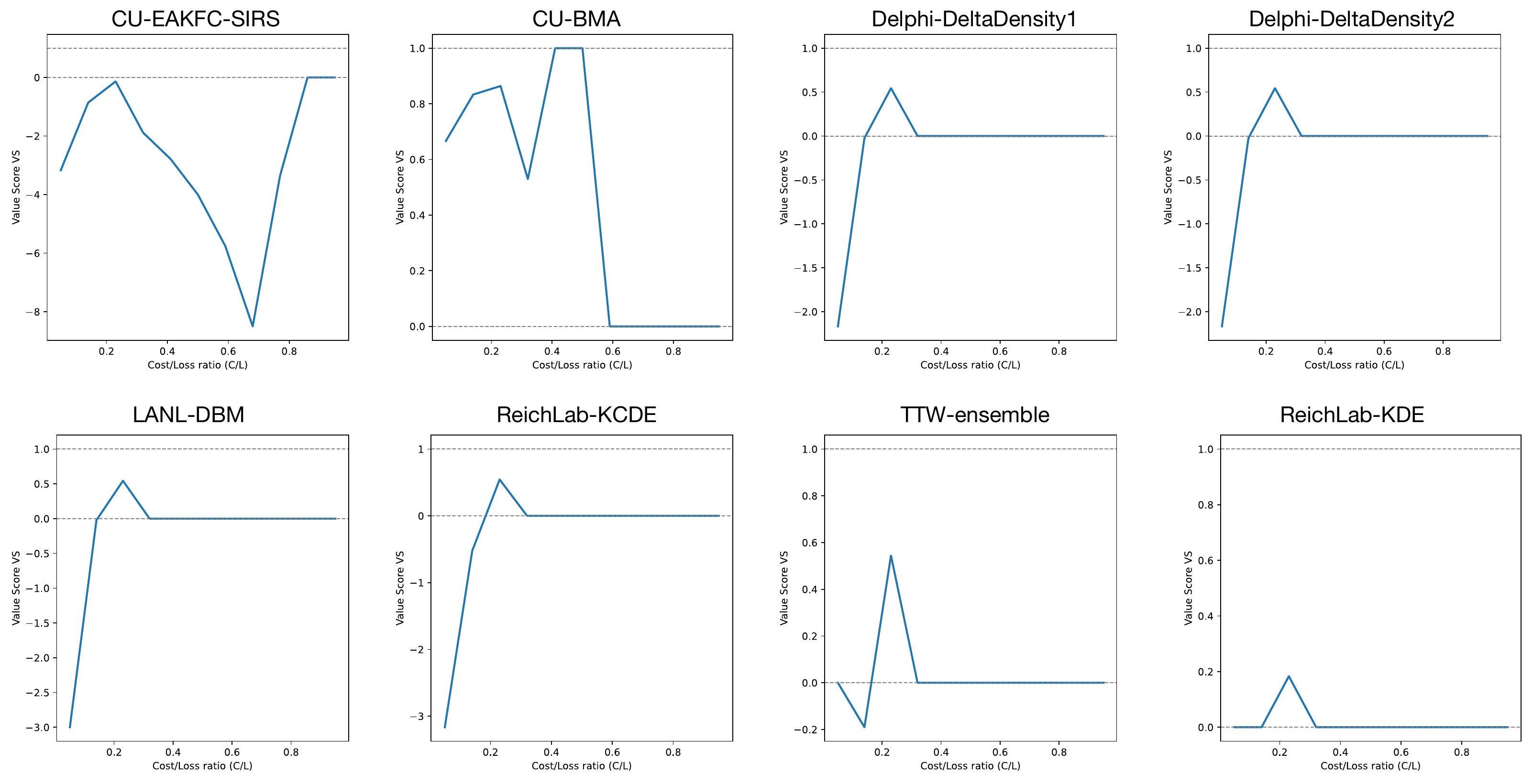}
\caption{\label{fig:overview}Value Score as a function of the cost-loss ratio $C/L$ for model forecasts of wILI peak intensity when the baseline event probability is calculated only on preceeding seasons.}
\end{figure}

\end{document}